**Accepted version, published in the *Journal of Financial Stability*, 2018, vol. 37, pp. 60–73, DOI:** https://doi.org/10.1016/j.jfs.2018.05.004.

Are Charter Value and Supervision Aligned? A Segmentation Analysis[*]

Juan Aparicio,
University Miguel Hernandez
(Spain)
Email: j.aparicio@umh.es

Miguel A. Duran,
University of Malaga
(Spain)
Email: maduran@uma.es

Ana Lozano-Vivas,
University of Malaga
(Spain)
Email: avivas@uma.es

Jesus T. Pastor,
University Miguel Hernandez
(Spain)
Email: jtpastor@umh.es

**Abstract:** Previous work suggests that the charter value hypothesis is theoretically grounded and empirically supported, but not universally. Accordingly, this paper aims to perform an analysis of the relations between charter value, risk taking, and supervision, taking into account the relations' complexity. Specifically, using the CAMELS rating system as a general framework for supervision, we study how charter value relates to risk and supervision by means of classification and regression tree analysis. The sample covers the period 2005–2016 and consists of listed banks in countries that were members of the Eurozone when it came into existence, along with Greece. To evaluate the crisis consequences, we also separately analyze four subperiods and countries that required financial aid from third parties and those that did not so, along with large and small banks. Our results reflect the complexity of the relations between charter value, supervision, and risk. Indeed, supervision and charter value seem aligned regarding only some types of risk.

---

[*] The authors gratefully acknowledge financial support from the Spanish Ministry of Economics and Competitiveness (Projects ECO2014-52345-P and MTM2016-79765-P (AEI/FEDER, UE)). We also thank the participants at the 10th International Risk Management Conference and the Sixth International Conference of the Financial Engineering & Banking Society for their helpful comments.

# 1. Introduction

A bank's charter value is the value of the right to continue to operate. Given that charter value is potentially lost when banks default, the so-called charter value hypothesis (CVH) has identified this value as a self-disciplinary device that disincentivizes excessive risk taking (Marcus 1984, Keeley 1990). Indeed, previous work has pointed out that charter value helps mitigate the moral hazard problem associated with the safety net, especially with deposit insurance, and helps align the interests of banks with their supervisors' (Demsetz et al. 1996). Thus, charter value can contribute to protecting debtholders and taxpayers from potential wealth expropriation caused by risk shifting (Jensen and Meckling 1976, Merton 1977). As Marshall and Prescott (2001: 142) put it, "high franchise banks are virtually self-regulating." By means of empirical tests in which a single type of risk is used in each econometric exercise, a number of papers have found empirical evidence supporting the CVH for the United States (Keeley 1990, Demsetz et al. 1996), Europe (Gropp and Vesala 2001), Japan (Konishi and Yasuda 2004), India (Ghosh 2009), and intercontinental sets of countries (González 2005, Haq et al. 2014).

Nevertheless, other theoretical and empirical contributions have challenged the CVH, showing that the relations between charter value, risk-taking behavior, and supervision/regulation can be less straightforward. Three main reasons lie behind the doubts about the universal validity of the CVH. First, bank risk and supervision are multidimensional; second, the risk–charter value relation may be circumstance dependent and, hence, subject to changes in time; and, third, nonlinearities may be present in such relation. In this sense, Park's (1997) model is based on the idea that there is a plurality of bank risks and supervision, as reflected by the CAMELS supervisory rating system, is concerned with different sources of risk. Accordingly, charter value is not necessarily related to all these sources in the same way under all circumstances. Indeed, a larger charter value can even lead banks to take higher risks unless supervision is tight. As Park shows, this result has a relevant policy implication: Charter value is not always a substitute for bank supervision. Consistently with the complexity of the relation between charter value and other bank features, Acharya (2003) shows that both the level of capital and regulatory forbearance buffer bank owners from the loss of charter value; therefore, charter value and bank capital can be inversely related if central banks adopt greater forbearance. From an empirical perspective and taking into account the polymorphic nature of risk, some studies have found evidence supporting that charter value is negatively related to idiosyncratic risk but positively related to systematic risk. Haq and Heaney (2012) find such a result for a sample of European banks. For the US banking system, Anderson and Fraser (2000) observe no relation between charter value and systematic risk.

Regarding the time variability of the relation between charter value and risk taking, this relation seems to change through the business cycle (Saunders and Wilson 2001). In a similar vein, different results across periods are found by Galloway et al. (1997), who report no statistical difference between the ex post risk-taking behavior of banks with low and high charter values over the period 1977–1982, whereas such differences are observed in 1983–1994. Finally, empirical evidence suggests a nonlinear relation between charter value and risk taking: As charter value increases, risk taking first decreases and then increases (Niu 2012). Nonlinearity is also observed by Jokipii (2009). Indeed, linearity seems at odds with the bipolar nature of the banking system that Marcus (1984) points out, with banks split between very safe and precarious institutions.

Therefore, previous work suggests that the CVH is theoretically grounded and empirically supported, but not universally, that is, not for any type of risk or level of charter value and regardless of supervisory and regulatory policy. This paper aims to analyze the relations between risk taking, supervision, and charter value using an empirical research strategy that takes into account the complexity of these relations and which can thus help shed additional light on this area. In particular, our strategy allows us to put at the core of the analysis characteristics of the risk–charter value relation almost completely disregarded by previous research. In this respect, following Park's (1997) model, our analysis is based on the principle that risk is multidimensional and the relations of charter value to those dimensions can be heterogeneous. Moreover, in tune with nonlinearities previously found by empirical research, our analysis takes into account the nonlinear, threshold-based nature of banking supervision's concerns

In a more detailed manner, our research strategy has two main features. First, to define the types of risk relevant to supervision, we use the aspects assessed under the CAMELS supervisory system: capital adequacy, asset quality, management capability, earnings quality, adequacy of liquidity, and sensitivity to market risk (Ioannides et al. 2010). Second, this multidimensional definition of risk under supervisory control is related to charter value by means of classification and regression trees (CART; see Breiman et al. 1984). In particular, this method allows us to find the CAMELS risk factors that better explain banks' charter value. CART operates by splitting the sample recursively into pairs of data subsets. This chain of splits results in a tree that starts at a root node, generates intermediate nodes, and ends at the terminal nodes, where no further meaningful splits are possible or a predefined stopping rule holds. In our analysis, sample banks are divided into two groups at any of these nodes by selecting the risk factor and its threshold most relevant to charter value, in the sense that those two groups have maximum internal homogeneity regarding charter value and are as distinct from each other as possible. Thus, our research strategy helps answer questions such as are charter value and supervision aligned regarding risk? For which types of risk are charter value and supervision aligned? What are the relevant thresholds of the variables capturing risk?

Although CART has been frequently used in other areas of economic research,[1] only a limited number of papers have applied it to the banking sector. Duttagupta and Cashin (2011) use CART to analyze banking crises in emerging and developing countries during 1990–2005. The same problem is examined by Manasse et al. (2013). Both Ioannides et al. (2010) and Alessi and Detken (2018) study early warning systems of banking crises. Emrouznejad and Anouze (2010) combine data envelopment analysis and CART to perform an efficiency analysis of the banking sector. Bijak and Thomas (2012) and Kao et al. (2012) apply CART to yet another area of banking research, customer credit scoring. These works reveal that CART can be particularly fruitful in some areas of the banking sector literature. One of these could be testing the CVH, since this method accommodates features of the relation between risk and charter value pointed out by previous work better than standard regression analysis does.

In this regard, the main advantages of CART for our analysis are as follows. i) Since multicollinearity is irrelevant in this method, we can include different types of bank risk in the analysis, despite their potential high correlation. Indeed, CART is a nonparametric statistical technique that does not rely on ordinary least squares assumptions or any other assumption about the probability distribution of the data. ii) Closely related to the previous point, instead of performing the analysis in terms of a single type of risk that can be statistically significant or not, the method itself singles out the main types of risk in relation to charter value. iii) A CART's binary division does not imply causality between a type of risk and charter value; it just means that one of the CAMELS risk factors splits the sample in a relevant way in relation to charter value. iv) The potential presence of nonlinearities in the relation between risk and charter value found by Jokipii (2009) and Niu (2012) does not affect the analysis that CART performs, since CART is itself a nonlinear method. v) Applying "tree-pruning" techniques has, among other consequences, that of cancelling potential effects of outliers, common in financial databases. vi) Finally, one of CART method's main assets is that it chooses not only the risk factors that split the sample in a relevant manner but also the thresholds at which those factors cause such splits. These threshold-based partitions are consistent with regulation and the modus operandi of supervisors.

---

[1] For instance, CART has been frequently applied in the analysis of financial crises. In this sense, Ghosh and Ghosh (2002) and Frankel and Wei (2004) focus on currency crises, whereas Manasse and Roubini (2009) and Savona and Vezzoli (2015) focus on sovereign debt crises.

Despite its advantages for the analysis of the supervision–risk–charter value relation, a common critique of CART is that it may lack robustness to modifications in the set of predictor variables or observations. To mitigate this drawback, we also use the random forest (RF) technique (Breiman 2001). This technique is a machine learning and data mining method that randomly selects predictors and portions of the sample to grow a multitude of trees. The forest thus obtained provides a measure of how important the explaining variables are in predicting the response variable, that is, it allows one to sort the predictor variable candidates and, hence, provides a criterion to select the best of them. Accordingly, as a preliminary stage, we obtain a 2,000-tree forest by applying the RF algorithm to alternative proxies of the CAMELS risk factors. Those proxies that score better in predicting charter value are then used in a second stage, as a set of robust predictors, to generate our CART-generated final tree. Similar approaches have been recently applied by Alessi and Detken (2017) to identify excessive credit growth and by Esteve et al. (2018) to classify tweets after a terrorist attack. Additionally, we use pruning techniques to control overfitting and thus obtain more reliable trees.

To analyze the relations between risk, supervision, and charter value, we use a sample that contains listed banks in the countries that were part of the Eurozone when it came into existence, along with Greece. The period studied extends from 2005 to 2016, although we also separately analyze the pre-crisis, financial crisis, sovereign debt crisis, and post-crisis periods. In addition, we independently analyze, on the one hand, those countries that required financial aid from third parties, that is, Portugal, Ireland, Greece, and Spain (PIGS), and the remainder of the countries and, on the other hand, large and small banks.

Our main results strengthen the conclusion that the relations between charter value, supervision, and risk are manifold. The trees that CART generates suggest that supervision and charter value are aligned in terms of only some sources of risk (earnings and liquidity risks), whereas they are misaligned regarding others (capital and systematic risks) and no evidence is observed for the remainder of the CAMELS risk factors (assets and management risks). Indeed, our tests of the CVH, not grounded on standard regression analysis, are consistent with the view that emerges from the review of previous research, which suggests that this hypothesis is not universally valid across types of risk. Such a limitation in the validity of the CVH has an important policy implication. It questions the prevalence of the principle that high-charter value banks are self-regulating and, hence, whether charter value can substitute supervision and regulation. In this regard, misalignments regarding capital risk and systematic risks imply that, even for banks with a high charter value, supervision seems especially required to control the adequacy of capital and systematically risky institutions. Such a conclusion is in tune with the strengthening of the capital requirements in Basel III. However, despite the new norms introduced for systemic risk, the new Basel Accord does not include the concrete regulation of systematic risk, even if the latter can induce the too-many-to-fail problem (Acharya and Yorulmazer 2007). This lack of regulation and the non-alignment of charter value and supervision in terms of systematic risk leave the entire control of this kind of risk in the hands of supervisors.

Another relevant finding refers to how the characteristics of the most heterogeneous subsamples generated by CART change in the post-crisis period. After the end of the crisis in the Eurozone, banks in the subsample with the highest mean charter value also have lower levels of risk for all the significant components of CAMELS.

This result suggests that the pressure that the crisis put on banks appears to have reinforced the self-disciplining role of charter value and, thus, eventually helped align charter value with supervision.

The analysis of non-PIGS and PIGS countries suggests that the relation between supervision and charter value is very different across the Eurozone. This heterogeneity implies that there is not a unique criterion shared by euro countries to define the potential areas in which, according to the CVH, charter value can substitute for supervision in risk control and, hence, supervision can be relaxed. Accordingly, any role of charter value in the common monitoring of the system under the Single Supervisory Mechanism should take into account regional differences in how this value could help control risk taking; otherwise, some risk factors could be insufficiently supervised in some areas, increasing the risk of asymmetric shocks (De Grauwe 2011).

In the analysis by bank size, we find that the main difference between large and small banks refers to liquidity risk; in particular, consistent with previous works that have pointed out that the latter are more financially constrained (Aschcraft et al. 2011), supervision and charter value are aligned for small banks and misaligned for large banks.

The remainder of the paper proceeds as follows. Section 2 explains the main features of the CART and RF methods. Section 3 describes the sample, the variables, and the use of the RF algorithm to select a set of sound predictors. Section 4 presents the results. Section 5 concludes the paper.

## 2. Methodological note on CART and RF

To examine how charter value relates to bank supervision and risk-taking behavior, we use CART.[2] This data mining technique is a nonparametric classification model introduced by Breiman at al. (1984).[3] The principle behind CART is relatively simple: A given criterion is chosen to recursively generate binary partitions of the data until no further meaningful division is possible or a stopping rule holds. The graphical result of this process is a tree (similar to the extensive-form representation of a game with perfect and complete information) that begins at the root node, develops through intermediate nodes, and ends at the terminal nodes or "leaves." The binary nature of CART is reflected in each (parent) node, except the leaves, giving birth to two child nodes. The split at any such root or intermediate node is caused by a predictor variable and a threshold for this variable. In data mining analysis a classification or regression tree is one in which the response variable to be analyzed is categorical or numerical, respectively. In this paper, we focus on the latter type of trees, due to the nature of our response variable.

---

[2] Specifically, we use the package rpart in R, by Therneu, Atkinson, and Ripley (https://cran.r-project.org/web/packages/rpart/index.html).
[3] See also Zhang and Singer (2010). For brief accounts of segmentation analysis and its application to economic problems, see, for example, Frankel and Wei (2004), Chamon et al. (2007), Manasse and Roubini (2009), Ioannides et al. (2010), Duttagupta and Cashin (2011), or Manasse et al. (2013).

Given all the possible ways to split the data at a node, CART builds regression trees by choosing those splits that maximally distinguish the response variable in the child nodes; that is, the algorithm recursively partitions the data into maximally heterogeneous groups. Before explaining in more detail how CART generates such partitions, let us introduce some nomenclature. At every node $t$, there are $n_t$ observations $(x_{1j}, x_{2j}, \dots, x_{mj}, y_j)$, where $x_{ij}$ is the value of predictor variable $i$ corresponding to observation $j$, with $i = 1,..,m$ and $j = 1,..,n_t$, and $y_j$ is the value of the response variable for $j$. The measure $\sum_{j \in t}(y_j - \bar{y}(t))^2$, with $\bar{y}(t) = \frac{1}{n_t}\sum_{j \in t} y_j$, represents the within sum of squares at node $t$, that is, it is the total squared deviation of the values of the response variable at this node with respect to their average.[4]

At any node $t$, CART implements a trial-and-error search based on trying multiple potential splits of the observations at $t$ and eventually choosing the best one. A split is defined as a pair $(x_i, u_i)$, where $u_i$ denotes a threshold for the $i$th predictor variable. If the pair $(x_{i'}, u_{i'})$ is selected by the algorithm as a potential split, the observations in $t$ are divided into $t_L^{(x_{i'}, u_{i'})} = \{j \in t: x_{i'} \leq u_{i'}\}$ and $t_R^{(x_{i'}, u_{i'})} = \{j \in t: x_{i'} > u_{i'}\}$; that is, the observations in $t$ are divided between a left child node, $t_L^{(x_{i'}, u_{i'})}$, that includes all the observations whose predictor variable $i'$ has a value not above the threshold $u_{i'}$, and a right one, $t_R^{(x_{i'}, u_{i'})}$, that includes the remainder of the observations. Among all potential splits, CART selects as the best one the pair $(x_i, u_i)$, such that $\max_{(x_i, u_i)} \{R(t) - R(t_L^{(x_i, u_i)}) - R(t_R^{(x_i, u_i)})\}$, where $R(t) = \frac{1}{n}\sum_{j \in t}(y_j - \bar{y}(t))^2$, that is, the chosen split is that which maximizes the dispersion of the response variable at node $t$ minus the dispersions of this variable at $t$'s child nodes. Thus, a regression tree is formed by iteratively splitting the sample to obtain child nodes that are as internally homogeneous as possible and, hence, as heterogeneous as possible.

---

[4] This sum of squares can also be interpreted as a measure of accuracy when the response variable $y$ is estimated by the mean $\bar{y}(t)$ of a set of observed values.

Once CART generates a split, the entire trial-and-error search is repeated for the resulting child nodes. The process continues until no further meaningful segmentations are possible or a predefined stopping rule is satisfied. Specifically, we require leaf nodes to have no fewer than 30 observations.

The predictive power of a CART-generated tree can be jeopardized by overfitting, that is, by including nodes excessively dependent on irrelevant features of the data, such as noise and outliers, or having too many predictor variables involved in the branching of the tree. One way to overcome this weakness is to prune the tree by removing the least reliable nodes. With this aim, we use the cross-validation model. This technique consists in randomly dividing the sample into $k$ subsets with as close as possible to the same number of observations. One of these subsets provides the testing data and the remainder the training data. The latter are used to build a tree, whereas the former test its accuracy. Then, the testing subset is rotated until the initial $k$ subsets have played this role. By minimizing cross-validated error, this training and testing process allows one to control anomalous or less consistent nodes and, hence, growing a non-overfitted, pruned tree.

Although CART performs very well in sample, a potential drawback is that it may not be robust as new predictor variables or observations are added to the analysis. To overcome this problem, we use the RF technique (Breiman 2001).[5] The RF technique generates multiple trees that are grown nondeterministically by means of a two-stage randomization procedure. In the first stage, random subsamples, bootstrapped from the original data set, are used to grow the trees, that is, each tree is based on a slightly different subsample. The second layer of randomization is provided by randomly selecting a subset of predictors at each node and using only these to split the sample. By applying the RF algorithm on a set of variables that proxy for the CAMELS risk factors in different ways, we obtain a ranking of potential predictors. Specifically, based on a 2,000-tree randomly generated forest, we can sort these proxies depending on their contribution to reducing the mean squared error (MSE) of the forest, that is, depending on the degree to which they contribute to increasing (decreasing) the explained (residual) variance of the response variable. Thus, by reducing both a model's bias and variance, RF allows us to select a robust set of explanatory variables capturing the CAMELS risk factors. We apply the CART technique to this set of variables to grow the tree that allows us to analyze the relations between charter value, supervision, and types of risk.

---

[5] Specifically, we use the package randomForest in R, by Breiman, Cutler, Liaw, and Wiener (https://cran.r-project.org/web/packages/randomForest/index.html).

The methodological features of CART make it particularly suitable for our research aim. As other nonparametric methods, CART is more effective when the data are ordinal or ranked. In this regard, following the supervisory authorities, we use a scale of one to five to rank the variables that capture the CAMELS risk factors. In addition, CART does not require the underlying assumptions of ordinary least squares regression, the most standard parametric method, to hold. In particular, CART is not affected by multicollinearity. This feature is relevant to our analysis, where the predictor variables are types of risk, which tend to be highly correlated. Another interesting characteristic of CART is that it does not require variables to be selected in advance: the CART algorithm itself identifies the most significant independent variables and eliminates nonsignificant ones. Therefore, CART allows us to analyze complex relations. Specifically, it allows us to find the types of risk that, at the thresholds indicated by the algorithm, split the sample into subsamples whose distributions of charter value are as heterogeneous as possible. Thus, for a type of risk that partitions the sample at a given threshold, we can determine whether the subsample with a higher average charter value is below or above the threshold. If that subsample is, for instance, below the threshold, we conclude that charter value and supervision are aligned in terms of this type of risk. Moreover, using thresholds to differentiate banks is similar to the way in which regulation usually defines risk-related requirements and supervision determines whether a financial institution is sound enough. The two last features of CART relevant to our research are that, on the one hand, CART can easily handle outliers, which are common in financial data sets, and, on the other hand, this technique is not affected by nonlinearities in the data, which seem to characterize the relation between charter value and risk (Jokipii 2009, Niu 2012).

## 3. Variables and data

Along with a description of the data set, this section defines the variables included in the analysis of the relations between charter value, risk, and supervision. In particular, we outline the set of variables measuring the CAMELS risk factors and discuss how they are selected based on the RF algorithm.

### 3.1. Data

Our data set consists of listed banks in the countries that, besides Greece, were members of the Eurozone when the euro was officially launched in 1999, specifically, Austria, Belgium, Finland, France, Germany, Ireland, Italy, the Netherlands, Portugal, and Spain. Since Luxembourg is an international hub for private banking, we exclude it to avoid potential distortions. The final sample covers the period 2005–2016 and includes 944 observations. The market value of bank equity is obtained from Datastream and Bankscope provides the data for the remainder of the variables. We manually merge the information extracted from these two databases.

To analyze separately the relations between charter value, risk, and supervision along the cycle, we split the sample into the periods before, during, and after the Great Recession. Following Truman (2013), we consider that the crisis in Europe started in 2008 and there are two distinct phases (see also Sá et al. 2016): The first one is from 2008 until the end of 2009, with the beginning of the crisis in Greece. The second phase, the so-called sovereign debt crisis, extends until 2013. In addition, we examine whether there are any differences in the manner in which charter value relates to risk and supervision, on the one hand, between the PIGS countries and the remainder of the sample countries, and, on the other hand, between large and small banking institutions.

### 3.2. Variables

The dependent variable of our empirical exercise is bank charter value. In tune with standard practice in empirical studies (e.g., Keeley 1990, Demsetz et al. 1996), charter value is captured by Tobin's Q, which is defined as

$$Q_{it} = \frac{MVE_{it} + BVL_{it}}{NTA_{it}},$$

where, for bank $i$ in period $t$, $MVE_{it}$ is the market value of equity, $BVL_{it}$ is the book value of liabilities, and $NTA_{it}$ is the book value of assets net of goodwill.

Regarding the variables used to measure bank risk, the Basel Committee on Banking Supervision (2015) indicates that most bank supervisory rating systems adopt a broad approach in which a range of indicators, rather than any single performance indicator, is used to determine a bank's financial position. Among such systems, the committee mentions the Uniform Financial Institutions Rating System, informally known as CAMELS,[6] as the standard. This rating system is also a reference for European authorities. In this sense, the vice president of the European Central Bank mentions in 2013 that the harmonization of the supervisory risk assessment frameworks in the Eurozone takes into account techniques to assess bank-specific risk indicators in the spirit of CAMELS.[7] In academic research, CAMELS is widely used as a standard way to identify the types of risk on which supervisors focus. For example, as part of the project to develop core conceptual frameworks supporting macroprudential supervision in the European Union, the Macroprudential Research Network uses CAMELS to identify vulnerabilities that can lead to distress in European banks (Betz et al. 2013).

Accordingly, we use CAMELS as a general framework that defines the different components of risk that supervision considers to assess the overall risk profile of banks. Thus, the analysis of the relations between these types of risk and bank charter value can shed light on whether charter value and risk relate to each other in a way that is consistent with supervision; specifically, we can observe the types of risk under the scrutiny of the supervisory authority, if any, that are associated with lower or higher charter values.

With the aim of making a robust selection of the variables capturing the CAMELS risk factors, we proceed in two stages. First, we define a set of proxies for these factors and use the RF algorithm to grow a 2,000-tree forest. Second, for each risk factor, we choose the proxy that contributes the most to increasing the explained variance of the response variable. The chosen variables feed the tree that we grow by means of CART.

---

[6] Introduced by US regulators in 1979, the acronym CAMELS refers to capital, asset, management, earnings, liquidity, and systematic risks. The latter, added in 1996, reflects banks' sensitivity to market risk.
[7] See https://www.ecb.europa.eu/press/key/date/2013/html/sp130212.en.html.

To define the proxies competing in the RF stage of the analysis, we first describe alternative variables capturing the CAMELS risk factors; then, we explain how these variables, following the CAMELS system practice, are rescaled using a continuous one-to-five scale. Indeed, the competing proxies of a risk factor could differ not only in the variables defining them but also in how these variables are rescaled.

Regarding the variables measuring the risk factors, capital adequacy is proxied by the capital ratio, that is, the ratio of equity to total assets, which is a standard measure of the risk of insolvency (e.g., Hughes and Mester 1993, Demsetz and Strahan 1997, Ioannides et al. 2010, Cole and White 2012).[8] In tune with the idea that loan loss allowances and provisions reflect credit portfolio quality and are required by bank supervisors to cover expected losses, we use the ratios of these allowances and provisions to total loans to capture asset risk (e.g., Benston and Wall 2005, Bikker and Metzemakers 2005, Ioannides et al. 2010, Cole and White 2012). To proxy for management capability, we compute, on the one hand, a measure of managers' ability to expand a bank's portfolio—the difference between the growth rates of the bank's loans and the annual nominal gross domestic product—and, on the other hand, two measures of managers' efficiency in generating income at lower cost—the ratio of non-interest expense to income and the ratio of non-interest expense to total assets (e.g., Ioannides et al. 2010, Igan and Pinheiro 2011, Haq et al. 2014, Bassett et al. 2015). Earnings risk is captured by the return on assets and the return on equity (e.g., Peek et al. 1999, Ioannides et al. 2010, Cole and White 2012, Bassett et al. 2015, Kupiec et al. 2017). In terms of liquidity, Hughes and Mester (1998) and Berlin and Mester (1999) indicate that the banking business produces risky illiquid loans while transforming them into safe liquid deposits, thus incurring liquidity risk. Accordingly, we capture this risk by the ratio of loans to deposits. In addition, we proxy for liquidity risk by the ratio of liquid assets to total assets (e.g., Peek et al. 1999, Ioannides et al. 2010, Bassett et al. 2015, Kupiec et al. 2017). Finally, systematic risk is measured by means of the beta coefficient, which, according to the capital asset pricing model, relates banks' stock returns and the market return. Returns are computed as the average of the differences between the logarithms of consecutive daily closing prices during a year and market prices are assumed to be given by Euro Stoxx 50 closing levels (e.g., Bakkar et al. 2017).

---

[8] Lack of data availability prevents us from using the capital adequacy ratio established by Basel III.

In relation to the rescaling of the ratios proxying for CAMELS risk factors, we use two different methods. The standard manner consists in assigning the scores between one and five proportionately to the real values of the proxies; thus, one (five) is the lowest (highest) level of risk and, for instance, if a variable is increasing in risk, two and three equal the first and second quartiles of the distribution of the proxy, respectively. Alternatively, under the assumption that the level of concern for supervisory authorities should not always follow a uniform distribution, scores of two or less are given only to institutions that satisfy a certain condition and, hence, are sound regarding the risk factor under assessment. Accordingly, following the criteria of the Uniform Financial Institutions Rating System,[9] we also rescale the risk-measuring variables differently. In particular, for a proxy that increases in risk, the scores in the closed interval $[1,2]$ and in the semi-open interval $(2,5]$ are proportionately distributed among the values of the proxy not above and above a given threshold, respectively. If the proxy were decreasing in risk, the scores would be assigned the other way around.

Given these two rescaling procedures, $Capt$ is the variable resulting from proportionately rescaling the equity ratio, whereas $Capt\_x$ stands for this ratio if the one-to-five scores are distributed using 6% as a threshold, that is, if the scores in $[1,2]$ and $(2,5]$ are distributed among banks with a capital ratio above and below this threshold, respectively. If rescaling is standard, $Asts$ and $Asts'$ stand for the rescaled ratios of loan loss allowances to loans and loan loss provisions to loans, respectively. If the scores between one and two are reserved for banks with a ratio of loan loss allowances (provisions) to loans below 1.5% (1%), the rescaled variable capturing asset risk is $Asts\_x$ ($Asts'\_x$). Under standard rescaling, $Mang$, $Mang'$, and $Mang''$ denote the three variables standing for management capability, that is, the gap between the growth rates of a bank's loans and the nominal gross domestic product, the ratio of non-interest expense to income, and the ratio of non-interest expense to total assets, respectively. If a threshold equal to 0.7 is used to rescale the ratio of non-interest expense to income, the rescaled variable is $Mang'\_x$. The two ratios measuring earnings risk, the return on assets and that on equity, become $Ergs$ and $Ergs'$, respectively, if they are rescaled in a standard way; however, if thresholds of 1% and 15% are used to rescale them, $Ergs\_x$ and $Ergs\_x'$ are the respective proxies of those ratios. With respect to liquidity risk, $Liqt$ and $Liqt\_x$ stand for the ratio of loans to deposits if rescaling is standard or based on a threshold equal to 0.8, respectively. The variable $Liqt'$ is the result of rescaling in a standard manner the ratio of liquid assets to total assets. Finally, regarding systematic risk, $Syst$ is equal to the beta coefficient rescaled in a standard way.

---

[9] See the Federal Deposit Insurance Corporation (2018).

3.3. RF-based variable selection and descriptive statistics

Figure 1 sorts the variables measuring the CAMELS risk factors. The sorting criterion is the percentage increase that removing one these proxies causes on the MSE of a 2,000-tree RF model.

**[Figure 1]**

Among the proxies, for instance, of asset risk, $Asts$ (in comparison to $Asts\_x$, $Asts'\_x$, and $Asts'$) is the one that reduces the MSE the most (24.5% vs. 23.4%, 22.7%, and 20.1%). For capital, management, earnings, and liquidity risks, the variables that satisfy this requirement of maximally reducing MSE are $Capt$ (46.6%), $Mang$ (21.6%), $Ergs\_x$ (42.6%), and $Liqt\_x$ (22.1%), respectively. According to these findings and taking into account the fact that $Syst$ captures systematic risk, we select $Capt$, $Asts$, $Mang$, $Ergs\_x$, $Liqt\_x$, and $Syst$ as the dependent variables of our CART analysis. For the sake of clarity, these six variables selected by the RF algorithm are called henceforth $C$, $A$, $M$, $E$, $L$, and $S$, respectively. Since such a selection is based on a process whose results are obtained by randomly removing variables and observations, we expect to minimize the effects of the potential instability associated with regression trees.

Since one (five) is the lowest (highest) level of risk in the one-to-five scale that we use to rescale risk factors, the higher (lower) the capital, management, and earnings (assets, liquidity, and systematic) risks, the lower the rescaled proxies of these risk factors selected by the RF algorithm, that is, $C$, $M$, and $E$ ($A$, $L$, and $S$). Table 1 displays the descriptive statistics of these six variables, without rescaling, and Tobin's Q for the whole sample, along with the subsamples of non-PIGS and PIGS countries.

**[Table 1]**

As Table 1 indicates, PIGS countries have a higher mean Tobin's Q (0.972 for non-PIGS vs. 1.014 for PIGS) but non-PIGS countries are, on average, less risky in four out of the six CAMELS risk factors; specifically, in assets (0.026 vs. 0.054), earnings (0.006 vs. 0.002), liquidity (0.889 vs. 1.043), and systematic (0.430 vs. 0.754) risks. Regarding capital risk, both sets of countries have the same capital ratio (0.072).

## 4. Results

This section presents the results from applying the CART method to the analysis of the relation between supervision and charter value. The first part of the section discusses the findings for the whole sample and the subperiods 2005–2007, 2008–2009, 2010–2013, and 2014–2016. The remaining two parts focus on the comparison between non-PIGS and PIGS countries and between large and small banks.

4.1. Charter value and supervision

According to the CVH, market value provides an alternative device for controlling bank risk (Marcus 1984, Keeley 1990) and, hence, tight supervision might not be as necessary for banks with high charter value. Nevertheless, previous findings have also pointed out that the CVH is not necessarily valid for all types of risk, at any level of risk, and regardless of the supervisory/regulatory framework; that is, charter value and supervision are not substitutes for each other under all circumstances (Park 1997). Our aim is to analyze whether charter value helps align the interests of supervisors and bank owners and, thus, contributes to mitigating the moral hazard problem in which banks shift risk to debtholders or, through the safety net, taxpayers (Jensen and Meckling 1976, Merton 1977). We consider that such alignment takes place if CART-based empirical evidence suggests that banks below a relevant threshold for that type of risk also have higher charter value. In tune with CAMELS, the types of risk of concern to supervisory authorities are capital, asset, management, earnings, liquidity, and systematic risks.

Segmentation analysis is used to find the types of risk for which charter value aligns with supervision. Figure 2 shows the trees that this method generates for the whole sample. At any given parent node, the left child node corresponds to banks that satisfy the inequality indicated at the parent node and the value of $Q$ shown at a terminal node is the average charter value of the banks that eventually reach this node.

**[Figure 2]**

In tune with the target of analyzing charter value-catalyzed alignments between shareholders' and supervisors' interests, we focus on the chains of splits that give rise to the terminal nodes with the most heterogeneous groups of banks, that is, those with the minimum and maximum average Tobin's Q, $Q^{Min}$ and $Q^{Max}$. Continuous lines indicate these chains in Figure 2. The lowest and highest mean charter values are 0.887 and 1.079, respectively. The path that leads to the former selects banks with a low $C$ (below 1.986) at the root node and, conditional on previous splits, a low $S$ (below 3.140), a high $L$ (above 2.446), and a low $C$ (below 1.650) at the rest of parent nodes. Along with the root node, $E$ is the only type of risk that divides the sample in the chain leading to the maximum value; specifically, conditional on having a high $C$ (above 1.986), banks with low $E$ (below 1.869) have the highest $Q$.[10]

The inequality that any node establishes for the corresponding splitting variable is "less than" in Figure 2. Hence, given a parent node, if the mean $Q$ of the group of banks on the left child node is higher than that on the right, the type of risk measured by the splitting variable at this parent node is positively related to charter value. When this is the case, banks and supervision would be aligned regarding this risk, unconditionally at the root node and conditionally otherwise. The first row in Table 2 (Eurozone) synthesizes the information conveyed by the tree in Figure 2 regarding whether charter value and supervision move in the same direction during the sample period. Based on this information, such an alignment is present just for $E$ and $L$ and conditionally, whereas misalignment is observed for $C$ and $S$ and no conclusions can be drawn for $A$ and $M$; that is, bank owners and supervisors do not seem to have aligned interests for most types of risk. Thus, from a methodological perspective not based on standard regression analysis, CART seems to provide additional support to previous works that have questioned the universal validity of CVH across types of risk, periods, or banking systems. The main policy implication of these findings is that banking authorities should be cautious about whether charter value is a self-discipline mechanism that helps control any risk and, thus, whether it can be considered an adequate substitute for the regulation and supervision of banking risk.

---

[10] Recall that $C$, $A$, $M$, $E$, $L$, and $S$ are rescaled variables and, hence, increasing in risk.

**[Table 2]**

Segmentation analysis can help understand the relations between risk, supervision, and charter value because it takes into account features such as the polymorphic nature of bank risk, the use of thresholds by supervisors and regulators, and the presence of nonlinearities and outliers. Nevertheless, we also check whether the findings based on CART are in line with a straightforward correlation analysis. With this aim, Table 3 includes the Pearson correlation coefficients between CAMELS risk factors and Tobin's Q. The comparison of Tables 2 and 3 indicates that, overall, there do not seem to be relevant inconsistencies between what CART trees and correlation coefficients suggest. In particular, consistently with the misalignment between supervision and charter value regarding $C$ and $S$, these two variables are positively correlated with $Q$ at the 1% significance level. Similarly, the alignment between supervision and charter value for $L$ is in tune with the significant and negative correlation coefficient between this proxy for liquidity risk and $Q$. The fact that $A$ is not a splitting variable in Figure 2 is also reflected by the correlation between $A$ and $Q$ being nonsignificant. The only differences that we find between Tables 2 and 3 do not imply a clear contradiction but we find a significant (insignificant) correlation where CART does not provide (provides) a splitting variable. This is the case for $M$ and $E$.

**[Table 3]**

Figure 3 shows the trees for the pre-crisis, financial crisis, sovereign debt crisis, and post-crisis periods. Focusing on the paths to $Q^{Min}$ and $Q^{Max}$, the comparison of these trees with that in Figure 3 suggests that the nature of the relations between charter value and risk are not altered in a major manner when the sample period is divided into subperiods; that is, the alignments or misalignments between supervision and banks regarding CAMELS risk factors do not change substantially through the phases of the crisis and after its end. In this sense, the partitions that $S$ and $C$ cause in the four subperiods, either at the root or intermediate nodes, are consistent with the misalignment between bank owners and supervisors, as in the entire sample period. However, as in the whole sample as well, these agents' interests are aligned regarding $E$ in all subperiods, except in 2010–2013, when this type of risk does not generate any segmentation. The main difference between the tree in Figure 2 and those in Figure 3 concerns $A$, which is not a splitting variable during the entire period but is so in 2010–2013.

**[Figure 3]**

In more detail, Figures 2 and 3 suggest, first, that $C$ plays a central role. For the tree of the entire sample, as for those of 2010–2013 and 2014–2016, capital risk is the splitting variable at the root node. In 2005–2007, it divides the sample at one of the two child nodes resulting from the root node. All these partitions suggest that $C$ is positively related to charter value. This positive relation between charter value and capital risk has been previously observed (Saunders and Wilson, 2001, Acharya 2003). It suggests that supervisors and banks are not aligned regarding capital risk: Bank owners have more to lose if the capital ratio is higher. Such an effect can be explained in terms of the equity multiplier: Given a bank's return on assets, the lower its equity, the higher the return on equity (Mishkin 2010); that is, bank shareholders prefer lower capital ratios because, ceteris paribus, the return on equity is higher.

The threshold at which $C$ splits the entire sample (1.986) implies that the mutually heterogeneous groups (in the sense of CART) with low and high Tobin's Q include 24.44% and 75.56% of the sample banks, respectively; that is, the high-$C$, high-$Q$ homogeneous group comprises about three-quarters of financial institutions. In 2010–2013 and 2014–2016, the thresholds at which $C$ splits the root nodes (1.686 and 2.063, respectively) imply that the percentage of banks in the high-$C$, high-$Q$ group is around 10 points higher during the sovereign debt crisis than during the post-crisis period, at 82.25% versus 73.26%, respectively. This difference reflects the fact that banks with an intermediate level of capital risk are now in the branch with a low $Q$ in 2014–2016. Accordingly, the end of the crisis seems to have aggravated the effects of the equity multiplier effect. The group of low-equity banks that have more to lose because their return on equity is higher has been restricted. After 2013, this group does not include financial institutions with intermediate levels of capital (specifically, with $C$ between 1.686 and 2.063).

The positive relation between charter value and capital risk is relevant not only to supervision but also to regulation. It suggests that banks with a high charter value are not those with a high capital ratio and, hence, it supports regulatory and supervisory measures aimed at guaranteeing an adequate amount of capital, even with a high charter value (Caprio and Summers 1993). In this sense, our result is in line with Basel III, which strengthens capital legal requirements.

Second, Figures 2 and 3 indicate that $S$ is also a key variable in the segmentation process. In the tree for the entire sample, it splits the child node following the root node in the path to the minimum $Q$. In the pre-crisis and financial crisis periods, systematic risk is the splitting variable at the root node. In these $S$-caused divisions, the banks in the path to the lowest $Q$ are those with a lower level of systematic risk. In the pre-crisis period, for example, banks with $S$ below 2.677 follow such path. Therefore, as $C$, $S$ seems to also be positively related to $Q$, so that charter value does not appear to be aligned with supervision regarding systematic risk either. This result is in tune with the work of Haq and Heaney (2012), who observe that charter value is positively related to systematic risk in the European banking system over the period 1996–2010.

In relation to this finding, a systematic bank is highly correlated with the market, that is, it is a financial institution that is likely to be in distress when the market is and, hence, a large number of other banks would also be in trouble. When this is the case, in tune with the too-many-to-fail principle, bailing out distressed banks can be ex post optimal for public authorities (Acharya and Yorulmazer 2007). Thus, as Demsetz et al. (1996) point out, banks with high systematic risk benefit from an additional, implicit government guarantee, which can help explain why these banks have a higher charter value. On this point, the threshold at which $S$ splits the 2005–2007 and 2008–2009 trees decreases from 2.677 to 2.262. This decrease suggests that banks may have had the perception that the implicit government guarantee spread, at least during the first phase of the European crisis, to most banks with an $S$ value above two, which is the score that marks the level of concern for supervisory authorities.

From a regulatory perspective, despite the new norms introduced for systemic risk, the new Basel Accord does not include the concrete regulation of systematic risk, even if the latter can induce the too-many-to-fail problem. This lack of regulation and the non-alignment of charter value and supervision in terms of systematic risk leave the entire control of this kind of risk in the hands of supervisors.

Third, our results suggest that $E$ is also a relevant variable in the segmentation analysis of how charter value relates to supervision. Earnings risk splits the right child node resulting from the root node in the tree for the entire sample and in the trees of the pre-crisis, financial crisis, and post-crisis periods. In these four cases, banks with a lower earnings risk score have a higher charter value. Therefore, conditional on the split at the root node, charter value and supervision seem aligned regarding this type of risk. Given that earnings risk is captured by the return on assets, this finding suggests that the group of banks that, on average, have more to lose if they go bankrupt (i.e., with the highest sample mean charter value) have higher current rates of return. In this respect, the transition from the pre-crisis period to the financial crisis brings about a significant change, which will be preserved during the post-crisis period. The threshold at which $E$ splits the sample increases from 1.635 up to levels close to three (2.951 and 2.845 in 2008–2009 and 2014–2016, respectively); that is, although charter value and supervision are aligned before the crisis just for banks with relatively low earnings risk, once the financial crisis starts, the threshold at which $E$ splits the sample rises to approximately the middle value of the one-to-five scale. Accordingly, the alignment between supervision and charter value regarding earnings risk seems to extend to a larger range of banks in the first period of the crisis and in the post-crisis, including banks in the middle of the profitability spectrum.

Fourth, in the path to the group of banks with a minimum average $Q$ in the tree of the entire sample, $L$ is also a splitting variable. Specifically, conditional on capital and systematic risks being low, banks that also have low liquidity risk have a higher mean charter value and, hence, charter value appears to self-discipline banks with a low $C$ and $S$ in relation to the liquidity risk they take. This result suggests that supervision and charter value are aligned in terms of this type of risk. Hence, although Basel III introduces a new global framework for liquidity regulation, supervisory concerns about this type of risk could be relaxed. Two caveats, however, qualify this conclusion. On the one hand, the fact that the negative relation between $L$ and $Q$ is found for banks with a low $C$ and $S$ implies that it applies to a restricted type of institutions, specifically, to 20.66% of sample banks. On the other hand, as Figure 3 shows, $L$ is not a splitting variable of the trees that result from dividing the sample into subperiods, at least in the paths to the lowest and highest $Q$ values. Out of these paths, $L$ splits the sample in the sovereign debt crisis.

Fifth, in the path to $Q^{Max}$ of the tree of the sovereign debt crisis, $A$ splits the group of banks with high capital risk. A similar result is observed in the post-crisis period, but not on the path to $Q^{Max}$. Since banks with higher asset risk have a higher mean $Q$, charter value and supervision are not aligned regarding this type of risk. Although this result puts into question the ability of charter value to control asset risk, we should take into account the fact that asset risk is measured by loan loss allowances over total loans. Therefore, high asset risk implies a high proportion of low-quality loans in the bank's portfolio, but it can also indicate that the bank is adequately protected against insolvency caused by those loans. In addition, the relevance of the positive relation between $A$ and $Q$ is called into question by the closeness of the splitting threshold of $A$ to the maximum value of the one-to-five scale (4.584). Indeed, only 9.97% of the banks with a high $C$ (above 1.686) in 2010–2013 are included in the homogeneous group of banks with a high $A$ and $Q^{Max}$.

To analyze in more detail the segmentation that CART generates, Table 4 shows the mean values of $Q$ and CAMELS risk factors at the terminal nodes of the paths to $Q^{Max}$ and $Q^{Min}$ in Figures 2 and 3. In contrast to these figures, the values displayed in Table 4 are not escalated. Given the two groups of banks with mean $Q^{Max}$ and $Q^{Min}$ in a tree, we use the Kolmogorov–Smirnov test to determine whether we can reject the null hypothesis that the distributions of $Q$ or the CAMELS factors in these groups are drawn from the same population. If the null hypothesis of the Kolmogorov–Smirnov test is rejected for a risk factor, we mark in gray the cell corresponding to the group of banks with the lowest level of risk.

**[Table 4]**

For the group of banks at the terminal nodes of the entire sample tree, the banks with the highest mean charter value (column (2) in Table 4) are less risky than the banks with the lowest mean charter value (column (1)) in $A$ (0.029 vs. 0.030), $M$ (0.661 vs. 0.450), $E$ (1.329 vs. 0.644), and $L$ (0.927 vs. 1.021), whereas the latter group is less risky in $C$ (0.130 vs. 0.061) and $S$ (0.200 vs. 0.670).[11] In 2005–2007, the least risky groups of banks across risk factors are the same as in the entire sample period. The results do not change in 2008–2009 for $M$, $E$, and $S$, but now the group of banks with the lowest mean $Q$ is less risky in $A$ and there do not appear to be significant differences between the banks with $Q^{Max}$ and $Q^{Min}$ regarding $C$ and $L$. During the sovereign debt crisis, the only variable for which there are no significant differences is $S$. For $C$, $A$, and $E$ ($M$ and $L$), the group of banks with $Q^{Min}$ ($Q^{Max}$) is less risky. After the end of the crisis, there is only one variable for which banks with $Q^{Min}$ are less risky: $C$, which is the splitting variable at the root node. On the contrary, banks with $Q^{Max}$ are less risky in $A$, $E$, and $L$.

---

[11] When the values of the variables are not escalated, a bank is less risky, all else being equal, if $C$, $M$, or $E$ increases or $A$, $L$, or $S$ decreases.

In relation to these results, recall that CART generates groups of banks that are as homogeneous as possible in terms of charter value. Accordingly, the results in Table 4 suggest that the relation between supervision and charter value is mixed, with no clear link between mean charter values in the homogeneous groups generated by CART and the mean values of the CAMELS risk factors. Banks with the highest mean charter value are less risky in some types of risk and riskier in others, but the types of risk for which this occurs are not constant in time. Nevertheless, after the crisis, the relation between risk and charter value becomes more straightforward: in 2014–2016, the group of banks with $Q^{Max}$ is less risky in all types of risk whose distributions are significantly different from those in the group of banks with $Q^{Min}$, except $C$. Therefore, once the Eurozone started coming out of the crisis, CART splits sample banks into two homogeneous groups. In the group with the highest charter value, this value and the risk factors for which the null of the Kolmogorov–Smirnov test is rejected move, on average, in the same direction, so that charter value and supervision seem to become aligned after the financial crisis. The only exception to such alignment is capital risk, which suggests that the equity multiplier effect is strong enough to make this risk the only source of misalignment between banks and supervision.

4.2. Non-PIGS and PIGS countries

We also repeat our analysis for two separate groups of countries. The criterion used to split the sample is based on whether countries received financial assistance from third parties either to finance their public debt or to rescue the banking system. Therefore, the first group includes Austria, Belgium, Finland, France, Germany, Italy, and the Netherlands, whereas the second comprises the PIGS countries of Greece, Ireland, Portugal, and Spain. Figures 4 and 5 show the trees for the non-PIGS and PIGS countries, respectively. Since no tree is generated for the latter countries in the subperiods if we maintain the stopping rule requiring at least 30 banks in the terminal nodes, this rule is not used to obtain the trees in panels (b) to (e) in Figure 5.

**[Figure 4]**

The second and third rows (non-PIGS and PIGS countries, respectively) in Table 2 are based on the paths to $Q^{Min}$ and $Q^{Max}$ of the trees in panel (a) of Figures 4 and 5, respectively. These rows show whether charter value and supervision are aligned for non-PIGS and PIGS countries in the sample period, whereas the first row (Eurozone) indicates the same for all sample countries. There seem to be no major contradictions between the alignments or misalignments in the entire set of sample countries and what we find in the non-PIGS and PIGS countries. In this sense, as in the whole sample, $Q$ is positively (negatively) related to $C$ ($L$) in non-PIGS countries and $Q$ is negatively related to $E$ in the PIGS sample. Therefore, the nature of the relations between supervision and charter value does not appear to change substantially in the samples of countries that received and did not receive financial assistance. The only contradiction is that the alignment between charter value and $Q$ regarding $E$ that we observe in the entire sample (and also in the PIGS countries) turns into misalignment in non-PIGS countries.[12]

Although the division into non-PIGS and PIGS countries does not appear to modify the nature of the relation between $Q$ and the types of risk, the comparison of Figures 4 and 5 indicates that the trees of these two sets of countries are very different. For PIGS countries, $E$ is the splitting variable at the root node of the tree for the sample period. Moreover, after the partition caused by $E$, no other variable subdivides the sample into homogeneous groups. The variable capturing earnings risk also splits the PIGS sample in the financial crisis and post-crisis periods. In all these splits, consistent with our results for the entire sample, supervision and charter value seem aligned, that is, banks with a better earnings risk score have a higher mean charter value. Therefore, the link between current profitability and having more to lose in case of bankruptcy seems to play a central role in the analysis of the alignment between charter value and supervision in PIGS countries.

**[Figure 5]**

---

[12] Another type of difference between our results for all the countries and the non-PIGS and PIGS countries in the entire period arises from segmentations present in only one of these samples. In this regard, $S$ and $A$ are splitting variables only in the entire sample and the non-PIGS sample, respectively.

For non-PIGS countries, by contrast, $E$ is only a splitting variable in the entire sample and conditional on banks having a low $C$ and $A$, that is, for 4.64% of non-PIGS banks. The central role in non-PIGS countries is played by $C$; it is the splitting variable at the root node in the entire sample period and in all subperiods, except 2005–2007, in which $C$ is also a splitting variable but on the left child node of the root node. Since these splits are all consistent with misalignment between supervision and banks regarding capital risk, the equity multiplier effect discussed for the entire sample seems especially strong in non-PIGS countries. The variable $C$ is also a splitting variable in PIGS countries (in 2008–2009, 2010–2013, and 2014–2016) but never at the root node.

The relation between charter value and supervision regarding $A$ is ambiguous in non-PIGS countries. This type of risk is a segmenting variable only when the entire sample period of these countries is analyzed, but it is so in two nodes: In one of them, $A$ and $C$ are negatively related and, in the other, positively. The role of $A$ seems more relevant and clearer in PIGS countries. Along with splitting a child node in the financial crisis, this type of risk is the splitting variable at the root node of the sovereign debt crisis. In both cases, supervision and charter value are misaligned. Such misalignment, also found for the entire sample, can be interpreted as the result of $A$ measuring asset risk by means of loan loss allowances and, hence, capturing how well provisioned banks are. Indeed, the ratio of loan loss allowances to loans in the PIGS countries increased exponentially during the crisis, from a trough of 0.023 in 2007 to a peak of 0.087 in 2013; furthermore, it rose at a much higher rate than in the non-PIGS countries.

Another difference between non-PIGS and PIGS countries is due to $L$. It is a splitting variable in the trees of non-PIGS countries corresponding to the sample period, the sovereign debt crisis, and the post-crisis period. In all these splits, as in the tree for all countries, charter value and supervision are aligned. Regarding PIGS countries, $L$ is only relevant in the 2014–2016 tree and not in the paths to $Q^{Max}$ or $Q^{Min}$. Therefore, the conclusion that, despite Basel III, regulatory and supervisory concerns on liquidity risk can be relaxed for high-charter banks seems to apply only to non-PIGS countries.

Accordingly, the comparison between non-PIGS and PIGS trees suggests that risk, supervision, and charter value relate to each other very differently across the Eurozone. This result highlights great differences that seem to exist between groups of European countries. Moreover, it suggests that, to make the banking system more resilient, supervisors cannot rely on charter value homogeneously across the euro area, unless further integration is previously achieved. Therefore, the measures adopted in the European Union to ensure consistent supervision should consider that the areas of special concern for supervision can differ across countries.[13] Otherwise, some types of risk in some countries could be inadequately supervised.

4.3. Large and small banks

A potential drawback of our analysis is that the results may be driven by a handful of large banks in the sample, given the relevance of size to bank soundness (Demsetz and Strahan 1997, Ioannides et al. 2010, Cole and White 2012, Kupiec et al. 2017). To check whether the nature of the relation between supervision and charter value is altered by bank size, we divide the sample into banks whose assets are below and above the sample median. Figure 6 shows the trees for both groups of banks and the last row in Table 2 indicates whether supervision and charter value are aligned for the CAMELS risk factors in these two groups. The nature of the relations between charter value and supervision regarding capital and earnings risk is the same for both large and small banks: $Q$ is positively related to $C$ and negatively related to $E$ in both subsamples. The only disparity between banks of different sizes relates to liquidity risk: Supervision and small banks are aligned regarding $L$, whereas misalignment is observed for large banks. In this respect, the trees in Figure 6 indicate that more valuable banks—that is, those with a higher charter value and, hence, more to lose in case of default—have lower liquidity risk only if they are small. This result can be explained by the different financial constraints that large and small banks face. The latter typically have greater access to funding sources and, especially, to the interbank credit markets. Thus, small banks are more valuable if they have less liquidity risk, but the opposite applies for large banks (Ashcraft et al. 2011).

**[Figure 6]**

---

[13] To promote the banking union, the two main initiatives of the European Union are the Single Supervisory Mechanism and the Single Resolution Mechanism.

## 5. Conclusion

Previous theoretical and empirical research supports the CVH, but not universally. This result underlies our paper's aim. As a contribution to the examination of this hypothesis, we analyze the relations between charter value, risk, and supervision, considering the multidimensional nature of bank risk. Specifically, we examine the relations between charter value and the six components assessed by the CAMELS rating system. If a low level of risk in any of these components is associated with a high charter value, we consider charter value and supervision to be aligned and, hence, in tune with the CVH, charter value can be considered a self-disciplining device regarding that component. We use CART to perform this analysis. Although it has been scarcely applied in the banking literature, the characteristics of CART can provide fruitful results in this area of research. In particular, it allows us to perform a threshold-based analysis of the relation between charter value and the set of supervised types of risk. To reinforce the robustness of the results, the variables capturing the CAMELS risk factors are selected by the RF method. Moreover, we prune trees to discard unreliable nodes and, hence, augment the predictive power of our final CART-generated trees.

Our results provide additional support for the idea that the relation between risk and charter value is complex, that is, the relation is not homogeneous, regardless of the type and level of risk or the period. In this sense, countering the idea that charter value and supervision are universal substitutes for each other, the findings in this paper suggest that charter value is self-regulating only for some types of risk, whereas, for others, tight supervision is required. Specifically, in the entire period studied, we observe that supervision and charter value are aligned regarding earnings and liquidity risks but misaligned for capital and systematic risk. The misalignment regarding capital risk supports the strengthening of capital regulations introduced by Basel III. Given the misalignment regarding systematic risk and the lack of regulation, controlling this risk factor and, thus, the too-many-to-fail problem depend on supervision.

Comparison of the trees of the non-PIGS and PIGS countries shows that the relation between risk and charter value in these areas is not homogeneous. Thus, the common supervisory framework of the Eurozone should consider that charter value can help control some risk factors, but unequally so across European regions. Finally, the tighter financial constraints of small banks help explain why the only difference that bank size causes in the alignment of supervision and charter value refers to liquidity risk.

**References**

Acharya, V. V. 2003. Is the International Convergence of Capital Adequacy Regulation Desirable. *Journal of Finance* 63: 2745-2781.

Acharya, V. V., and T. Yorulmazer. 2007. Too Many to Fail—An Analysis of Time Inconsistency in Bank Closure Policies. *Journal of Financial Intermediation* 16: 1-31.

Alessi, L., and C. Detken. 2018. Identifying excessive credit growth and leverage. *Journal of Financial Stability*. http://dx.doi.org/10.1016/j.jfs.2017.06.005.

Anderson, R. C., and D. R. Fraser. 2000. Corporate Control, Bank Risk Taking, and the Health of the Banking Industry. *Journal of Banking and Finance* 24: 1383-1398.

Ashcraft, A., J. McAndrews, and D. Skeie. 2011. Precautionary Reserves and the Interbank Market. *Journal of Money, Credit and Banking* 43: 311-348.

Bakkar, Y., C. Rugemintwari, and A. Tarazi. 2017. Charter Value and Bank Stability Before and After the Global Financial Crisis of 2007–2008. Mimeo,

Basel Committee on Banking Supervision. 2015. Report on the Impact and Accountability of Banking Supervision. Bank of International Settlements.

Bassett, W. F. , S. J. Lee, and T. W. Spiller. 2015. Estimating Changes in Supervisory Standards and Their Economic Effects. *Journal of Banking & Finance* 60: 21-43.

Benston, G. J., and L. D. Wall. 2005. How Should Banks Account for Loan Losses, *Journal of Accounting and Public Policy* 24: 81-100.

Berlin, M., and L. J. Mester. 1999. Deposits and Relationship Lending. *Review of Financial Studies* 12: 579-607.

Betz, F., S. Oprica, T. O. Peltonen, and P. Sarlin. 2013. Predicting Distress in European Banks. ECB Working Paper Series, nº. 1,597.

Bijak, K., and L. C. Thomas. 2012. Does segmentation always improve model performance in credit scoring? *Expert Systems with Applications* 39: 2433-2442.

Bikker, J. A., and P. A. J. Metzemakers. 2005. Bank Provisioning Behavior and Procyclicality. *Journal of International Financial Markets, Institutions and Mon*ey 15: 141-157.

Breiman, L. 2001. Random Forest. *Machine Learning* 45: 5-32.

Breiman, L., J. Friedman, R. A. Olshen, and C. J. Stone. 1984. *Classification and Regression Trees*. London: Chapman and Hall.

Caprio Jr., G., and L. H. Summers. 1993. Finance and Its Reform. Beyond Laissez-Faire. The World Bank Working Papers, nº. 1,171.

Chamon, M., P. Manasse, and A. Prati. 2007. Can We Predict the Next Capital Account Crisis? IMF Staff Papers 54: 270-305.

Cole, R. A, and L. J. White. 2012. Déjà Vu All Over Again: The Causes of US Commercial Bank Failures *This* Time Around. *Journal of Financial Services Research* 42: 5-29.

De Grauwe, P. 2011. The Governance of A Fragile Eurozone. CEPS Working Document, n.º 346.

Demsetz, R. S., M. R. Saindenberg, and P. E. Strahan. 1996. Banks with Something to Lose: The Disciplinary Role of Franchise Value. *FRBNY Economic Policy Review* October: 1-14.

Demsetz, R. S., and P. E. Strahan. 1997. Diversification, Size, and Risk at Bank Holding Companies. *Journal of Money, Credit and Banking* 29: 300-313.

Duttagupta, R., and P. Cashin. 2011. Anatomy of Banking Crises in Emerging and Developing Market Countries. *Journal of International Money and Finance* 30: 354-76.

Emrouznejad, A., and A. L. Anouze. 2010. Data Envelopment Analysis with Classification and Regression Tree—A Case of Banking Efficiency. *Expert Systems* 27: 231-246.

Esteve, M., F. Miro, and A. Rabasa. 2018. Classification of Tweets with a Mixed Method Based on Pragmatic Content and Meta-Information. *International Journal of Design & Nature and Ecodynamics* 13: 60-70.

Federal Deposit Insurance Corporation. 2018. Risk Management Manual of Examination Policies. https://www.fdic.gov/regulations/safety/manual/.

Frankel, J. A., and S. Wei. 2004. Managing Macroeconomic Crises. NBER Working Papers, nº. 10,907.

Galloway, T. M., B. L. Winson, and D. M. Roden. 1997. Banks' Changing Incentives and Opportunities for Risk Taking. *Journal of Banking and Finance* 21: 509-527.

Ghosh, S. 2009. Charter Value and Risk-Taking: Evidence form Indian Banks. *Journal of the Asian-Pacific Economy* 14: 270-286.

Ghosh, S. R., and A. Ghosh. 2002. Structural Vulnerabilities and Currency Crises. I*MF Staff Papers*, 50: 481-506.

González, F. 2005. Bank Regulation and Risk-Taking Incentives: An International Comparison of Bank Risk. *Journal of Banking and Finance* 29: 1153-1184.

Gropp, R., and J. Vesala. 2001. Deposit Insurance and Moral Hazard: Does the Counterfactual Matter? ECB Working Paper Series, nº. 47.

Haq, M, and R. Heaney. 2012. Factors Determining European Bank Risk. *Journal of International Financial Markets, Institutions and Money* 22: 696-718.

Haq, M., R. Faff, R. Seth, and S. Mohanty. 2014. Disciplinary Tools and Bank Risk Exposure. *Pacific-Basin Finance Journal* 26: 37-64.

Hughes, J. P., and L. J. Mester. 1993. A quality and risk-adjusted cost function for banks: Evidence on the "To-Big-To-Fail" Doctrine. *Journal of Productivity Analysis* 4: 293-315.

Hughes, J. P., and L. J. Mester. 1998. Bank Capitalization and Cost: Evidence of Scale Economies in Risk Management and Signaling, *Review of Economics and Statistics* 80: 314-325.

Igan, D. O., and M. Pinheiro. 2011. Credit Growth and Bank Soundness: Fast and Furious. IMF Working Paper, WP/11/278.

Ioannides, C., F. Pasiouras, and C. Zopounidis. 2010. Assessing Bank Soundness with Classification Techniques. *Omega* 38: 345-357.

Jensen, M. C., and W. H. Meckling. 1976. Theory of the Firm: Managerial Behavior, Agency Costs, and Capital Structure. *Journal of Financial Economics* 3, 305-360.

Jokipii, T. 2009. Nonlinearity of Bank Capital and Charter Values. Mimeo.

Kao, L. J., C. C. Chiu, and F. Y. Chiu. 2012. A Bayesian latent variable model with classification and regression tree approach for behavior and credit scoring. *Knowledge-Based Systems* 36: 245-252.

Keeley, M. C. 1990. Deposit Insurance, Risk, and Market Power in Banking. *American Economic Review* 80: 1183-1200.

Konishi, M., and Y. Yasuda. 2004. Factors Affecting Bank Risk-Taking: Evidence from Japan. *Journal of Banking and Finance* 28: 215-234.

Kupiec, P., Y. Lee, and C. Rosenfeld. 2017. Does Bank Supervision Impact Bank Loan Growth. *Journal of Financial Stability* 28: 29-48.

Manasse, P., and N. Roubini. 2009. "Rules of Thumb" for Sovereign Debt Crises. *Journal of International Economics* 78: 192-205.

Manasse, P., R. Savona, and M. Vezzoli. 2013. Rules of Thumb for Banking Crises in Emerging Markets. Università di Bologna DSE Working Paper, nº. 872.

Marcus, A. J. 1984. Deregulation and Bank Financial Policy. *Journal of Banking and Finance* 8: 557-565.

Marshall, D. A., and E. S. Prescott. 2001. Bank Capital Regulation with and without State-Contingent Penalties. *Carnegie-Rochester Conference Series on Public Policy* 54: 139-184.

Merton, R. C. 1977. An Analytic Derivation of the Cost of Deposit Insurance and Loan Guarantees. *Journal of Banking and Finance* 1: 3-11.

Mishkin, F. S. 2010. The Economics of Money, Banking & Financial Markets. Addison-Wesley, New York. 9th edition.

Niu, J. 2012. An Empirical Analysis of the Relationship between Charter Value and Risk Taking. *Quarterly Review of Economics and Finance* 52: 298-304.

Park, S. 1997. Risk-Taking Behavior of Banks under Regulation. *Journal of Banking and Finance* 21: 491-507.

Peek, J., E. S. Rosengren, and G. M. B. Tootell. 1999. Is Bank Supervision Central to Central Banking? *The Quarterly Journal of Economics* 114: 629-653.

Sá, T. M., E. D. Neves, and C. G. Góis. 2016. The Influence of Corporate Governance on Changes in Risk Following the Global Financial Crisis: Evidence from the Portuguese Stock Market. *Journal of Management & Governance*. D.O.I.:10.1007/s10997-016-9361-5.

Saunders, A., and B. Wilson. 2001. An Analysis of Bank Charter Value and Its Risk-Constraining Incentives. *Journal of Financial Services Research* 19: 185-195.

Savona, R., and M. Vezzoli. 2015. Fitting and Forecasting Sovereign Defaults using Multiple Risk Signals. *Oxford Bulletin of Economics and Statistics* 77: 66-92.

Truman, E. M. 2013. Asian and European Financial Crises Compared. PIIE Working Paper, nº. 13-9.

Zhang, H., and Singer, B. 2010. *Recursive Partitioning and Applications*. New York: Springer.

## Table 1. CAMELS risk factors: Summary statistics

This table reports descriptive statistics of Tobin's Q, $Q$, and the variables selected by the RF algorithm as proxies for the CAMELS risk factors $C$, $A$, $M$, $E$, $L$, and $S$, which stand for capital, asset, management, earnings, liquidity, earnings, and systematic risks, respectively. The values in this table are not rescaled. The top panel refers to the entire sample, whereas the other two refer to non-PIGS and PIGS countries, respectively.

| Risk factors | Obs. | Mean | Std. dev. | Min. | Max. |
|---|---|---|---|---|---|
| **Eurozone** | | | | | |
| $Q$ | 944 | 0.991 | 0.077 | 0.695 | 1.588 |
| $C$ | 944 | 0.072 | 0.042 | 0.048 | 0.257 |
| $A$ | 944 | 0.039 | 0.047 | 0.007 | 0.172 |
| $M$ | 944 | 1.223 | 1.301 | -2.314 | 8.401 |
| $E$ | 944 | 0.004 | 0.016 | -0.206 | 0.073 |
| $L$ | 944 | 0.958 | 0.339 | 0.029 | 3.865 |
| $S$ | 944 | 0.575 | 0.520 | -3.383 | 2.465 |
| **Non-PIGS countries** | | | | | |
| $Q$ | 520 | 0.972 | 0.069 | 0.695 | 1.463 |
| $C$ | 520 | 0.072 | 0.041 | 0.004 | 0.357 |
| $A$ | 520 | 0.026 | 0.023 | 0.011 | 0.112 |
| $M$ | 520 | 1.133 | 2.557 | -2.069 | 8.061 |
| $E$ | 520 | 0.006 | 0.009 | -0.072 | 0.040 |
| $L$ | 520 | 0.889 | 0.359 | 0.029 | 3.865 |
| $S$ | 520 | 0.430 | 0.472 | -0.309 | 2.015 |
| **PIGS countries** | | | | | |
| $Q$ | 424 | 1.014 | 0.079 | 0.870 | 1.588 |
| $C$ | 424 | 0.072 | 0.043 | -0.048 | 0.354 |
| $A$ | 424 | 0.054 | 0.063 | 0.007 | 0.172 |
| $M$ | 424 | 1.498 | 2.892 | -2.314 | 8.401 |
| $E$ | 424 | 0.002 | 0.021 | -0.206 | 0.073 |
| $L$ | 424 | 1.043 | 0.292 | 0.113 | 2.111 |
| $S$ | 424 | 0.754 | 0.522 | -3.383 | 2.465 |

### Table 2. Alignment and misalignment of charter value and supervision

This table shows whether the splits that the risk-measuring variables generate are consistent with the alignment or misalignment of charter value and supervision. Yes (no) means that they are aligned (misaligned) regarding the type of risk indicated in the risks factors columns. The CART trees to which this table refers are those of the whole sample (Figure 2); the subsamples of non-PIGS and PIGS countries (panel (a) of Figures 3 and 4); and the subsamples of small and large banks (Figure 6). We focus on the paths to the minimum and maximum average Tobin's Q. The variables $C$, $A$, $M$, $E$, $L$, and $S$ represent the rescaled proxies selected by the RF algorithm to capture capital, asset, management, earnings, liquidity, earnings, and systematic risks, respectively.

| Sample | Risk factors | | | | | |
|---|---|---|---|---|---|---|
| | $C$ | $A$ | $M$ | $E$ | $L$ | $S$ |
| **Eurozone** | No | - | - | Yes | Yes | No |

| Sample | $C$ | $A$ | $M$ | $E$ | $L$ | $S$ |
|---|---|---|---|---|---|---|
| **Non-PIGS** | No | Ambig[a] | - | No | Yes | - |
| **PIGS** | - | - | - | Yes | - | - |

| Sample | $C$ | $A$ | $M$ | $E$ | $L$ | $S$ |
|---|---|---|---|---|---|---|
| **Small banks** | No | - | - | Yes | Yes | - |
| **Large banks** | No | - | - | Yes | No | - |

### Table 3. Correlations between risk factors and Tobin's Q

This table shows the Pearson correlation coefficients between Tobin's Q and the CAMELS risk factors. The *p*-values are displayed in parentheses. Coefficients significant at 10% are in bold. The variables $C$, $A$, $M$, $E$, $L$, and $S$ represent the rescaled proxies selected by the RF algorithm to capture capital, asset, management, earnings, liquidity, earnings, and systematic risks, respectively, and $Q$ stands for Tobin's Q.

| | Risk factors | | | | | |
|---|---|---|---|---|---|---|
| | $C$ | $A$ | $M$ | $E$ | $L$ | $S$ |
| $Q$ | **0.326** (<.0001) | -0.017 (0.5981) | **-0.137** (<.0001) | -0.011 (0.7455) | **-0.121** (0.0002) | **0.279** (<.0001) |

---

[a] The tree of the non-PIGS countries in the entire sample period includes two nodes where $A$ is the splitting variable. These lead us to contradictory conclusions about the alignment of $Q$ and supervision regarding this type of risk.

## Table 4. CAMELS risk factors and charter value: Mean analysis

The columns in this table correspond to the paths to the maximum or minimum Tobin's Q of the trees in Figures 2 and 3, that is, the trees generated by CART for the entire sample and 2005–2007, 2008–2009, 2010–2013, and 2014–2016. The paths row shows the inequalities that split the samples along these paths, with the superscript $R$ indicating root nodes. The remaining rows show the mean values of the charter value and the CAMELS risk factors at the terminal nodes of the paths. The variable $Q$ stands for charter value, whereas $C$, $A$, $M$, $E$, $L$, and $S$ represent the proxies selected by the RF algorithm to capture capital, asset, management, earnings, liquidity, and systematic risks, respectively. In contrast to how results are shown in Figures 1 to 6, the variables are not escalated. Regarding the two groups of banks with mean maximum and minimum $Q$ values at the terminal nodes of a given tree, first, we use the Kolmogorov–Smirnov test to check whether we can reject the null that the distributions of any of our variables in these groups are drawn from the same population. We indicate whether the null can be rejected in the first of the two columns corresponding to the tree. Statistical significance at the 1%, 5%, and 10% levels is denoted by ***, **, and *, respectively. Second, a black cell highlights which of the two groups of banks has the maximum mean $Q$ and a gray cell indicates the group with the lowest level of risk of the CAMELS risk factors reported in that row if the null of the Kolmogorov–Smirnov test for this factor is rejected.

|   | Sample period 2005–2016 | | Pre-crisis period 2005–2007 | | Financial crisis 2008–2009 | | Sovereign debt crisis 2010–2013 | | Post-crisis period 2014–2016 | |
|---|---|---|---|---|---|---|---|---|---|---|
|   | (1) | (2) | (3) | (4) | (5) | (6) | (7) | (8) | (9) | (10) |
| **Paths** | $C^R > 0.091$ $S < 0.571$ $L < 0.898$ $C > 0.010$ | $C^R < 0.091$ $E > 0.960$ | $S^R < 0.206$ $C > 0.096$ | $S^R > 0.206$ $E > 1.103$ | $S^R < 0.097$ | $S^R > 0.097$ $E > 0.670$ | $C^R > 0.112$ | $C^R < 0.112$ $A > 0.049$ | $C^R > 0.088$ | $C^R < 0.088$ $E > 0.843$ |
| $Q$ | 0.887*** | 1.079 | 0.889*** | 1.134 | 0.957*** | 1.047 | 0.906*** | 0.9896 | 0.906*** | 1.027 |
| $C$ | 0.130*** | 0.061 | 0.152*** | 0.086 | 0.074 | 0.070 | 0.131*** | 0.055 | 0.130*** | 0.071 |
| $A$ | 0.030*** | 0.029 | 0.025** | 0.023 | 0.109* | 0.204 | 0.031*** | 0.046 | 0.059** | 0.039 |
| $M$ | 0.450*** | 0.661 | 0.248*** | 2.526 | 4.576** | 7.777 | -0.125*** | 4.104 | -0.628 | -0.909 |
| $E$ | 0.644*** | 1.329 | 1.509** | 2.118 | 0.434*** | 1.128 | 1.076*** | -0.299 | 0.527 | 0.904 |
| $L$ | 1.021*** | 0.927 | 1.026*** | 0.954 | 1.032 | 1.066 | 0.964*** | 0.954 | 0.911*** | 0.833 |
| $S$ | 0.200*** | 0.670 | 0.283*** | 0.879 | 0.213*** | 0.681 | 0.223 | 0.708 | 0.257 | 0.188 |

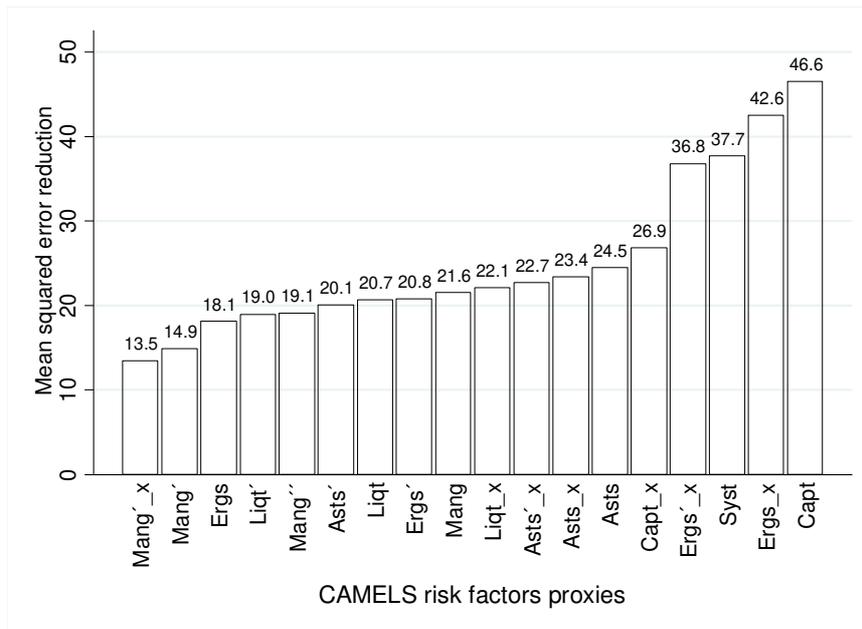

**Fig. 1. Random forest.** This figure shows the contribution of the rescaled proxies of the CAMELS risk factors to reducing the MSE of a 2,000-tree forest generated by the RF algorithm. The reductions are shown above the bars in percentage points.

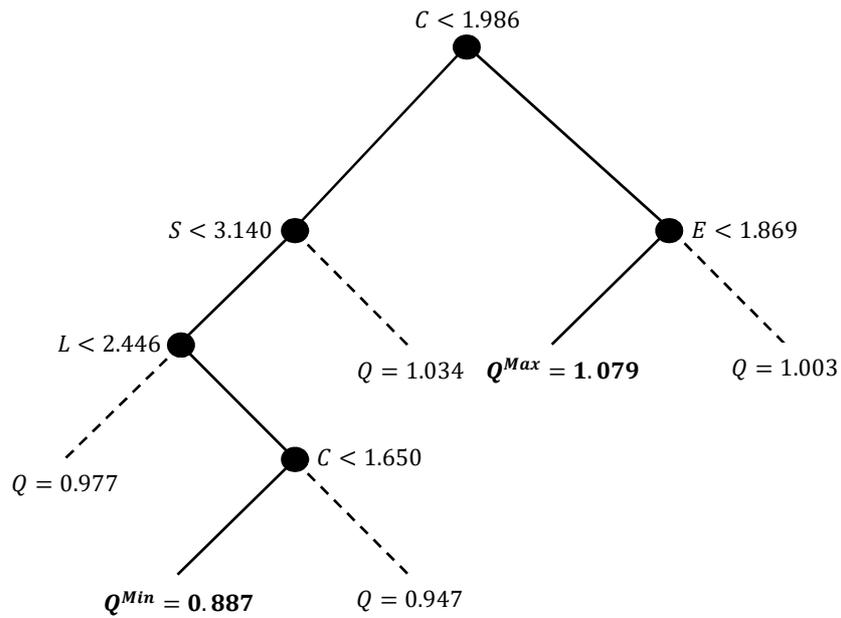

**Fig. 2. Supervision, risk, and charter value.** This figure shows the tree generated by CART for the total sample in 2005–2016. The variables $C$, $E$, $L$, and $S$ represent the proxies selected by the RF algorithm to capture capital, earnings, liquidity, earnings, and systematic risk, respectively, and $Q$ measures charter value. The variables capturing types of risk are rescaled using a scale of one to five. At any given node, the left branch includes banks satisfying the inequality shown in that node. The value of $Q$ at a terminal node equals the mean charter value of the group of banks in that terminal node. $Q^{Min}$ and $Q^{Max}$ stand for the minimum and maximum, respectively, of all the mean charter values corresponding to the sets of banks at the terminal nodes.

| Insert Fig. 3.a (see below) | Insert Fig. 3.b (see below) |
|---|---|
| **(a) Pre-crisis period, 2005–2007** | **(b) Financial crisis, 2008–2009** |
| Insert Fig. 3.c (see below) | Insert Fig. 3.d (see below) |
| **(c) Sovereign debt crisis, 2010–2013** | **(d) Post-crisis, 2014–2016** |

**Fig. 3. Supervision, risk, and charter value, by subperiod.** This set of figures show the trees generated by CART for the pre-crisis, financial crisis, sovereign debt crises, and post-crisis periods. The variables $C$, $A$, $E$, $L$, and $S$ represent the proxies selected by the RF algorithm to capture capital, credit, earnings, liquidity, and systematic risk, respectively, and $Q$ measures charter value. The variables capturing types of risk are rescaled using a scale of one to five. At any given node, the left branch includes banks satisfying the inequality shown in that node. The value of $Q$ at a terminal node equals the mean charter value of the group of banks in that terminal node. $Q^{Min}$ and $Q^{Max}$ stand for the minimum and maximum, respectively, of all the mean charter values corresponding to the sets of banks at the terminal nodes.

Fig. 3.a

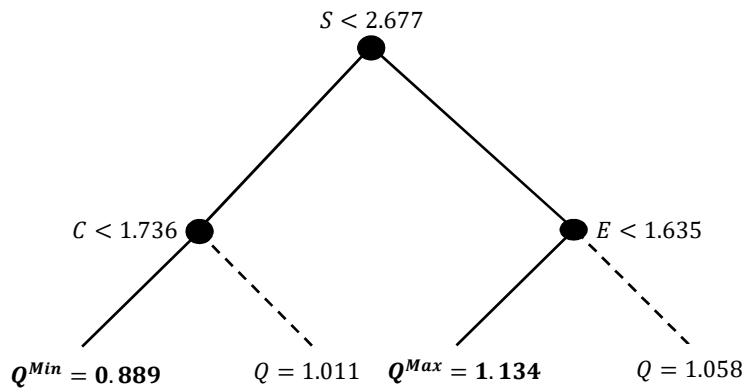

Fig. 3.b

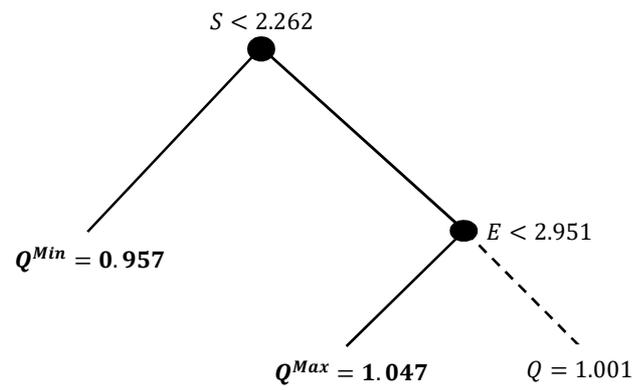

Fig. 3.c

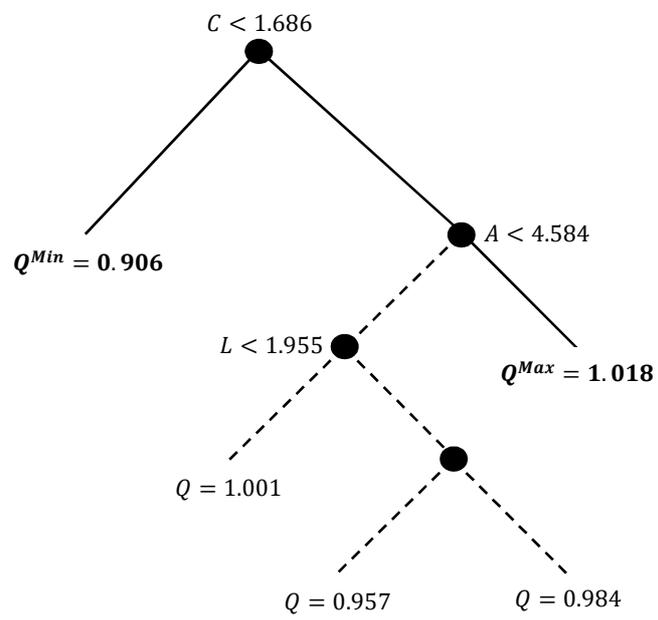

Fig. 3.d

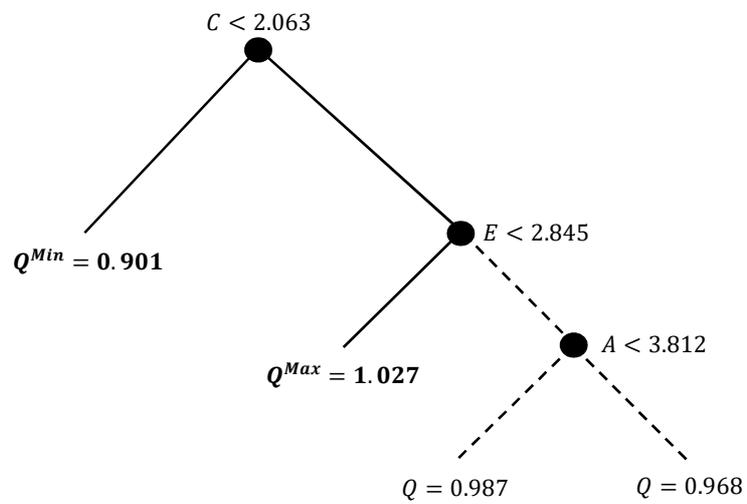

Insert Fig. 4.a
(see below)

**(a) Sample period, 2005–2016**

Insert Fig. 4.b          Insert Fig. 4.c
(see below)              (see below)

**(b) Pre-crisis period, 2005–2007**        **(c) Financial crisis, 2008–2009**

Insert Fig. 4.d          Insert Fig. 4.e
(see below)              (see below)

**(d) Sovereign debt crisis, 2010–2013**    **(e) Post-crisis, 2014–2016**

**Fig. 4. Non-PIGS countries.** This set of figures show the trees generated by CART for the non-PIGS countries in 2005–2016 and the pre-crisis, financial crisis, sovereign debt crises, and post-crisis periods. The variables $C, A, E, L$, and $S$ represent the proxies selected by the RF algorithm to capture capital, asset, earnings, liquidity, and systematic risks, respectively, and $Q$ measures charter value. The variables capturing types of risk are rescaled using a scale of one to five. At any given node, the left branch includes banks satisfying the inequality shown in this node. The value of $Q$ at a terminal node equals the mean charter value of the group of banks in that terminal node. $Q^{Min}$ and $Q^{Max}$ stand for the minimum and maximum, respectively, of all the mean charter values corresponding to the sets of banks at the terminal nodes.

Fig. 4.a

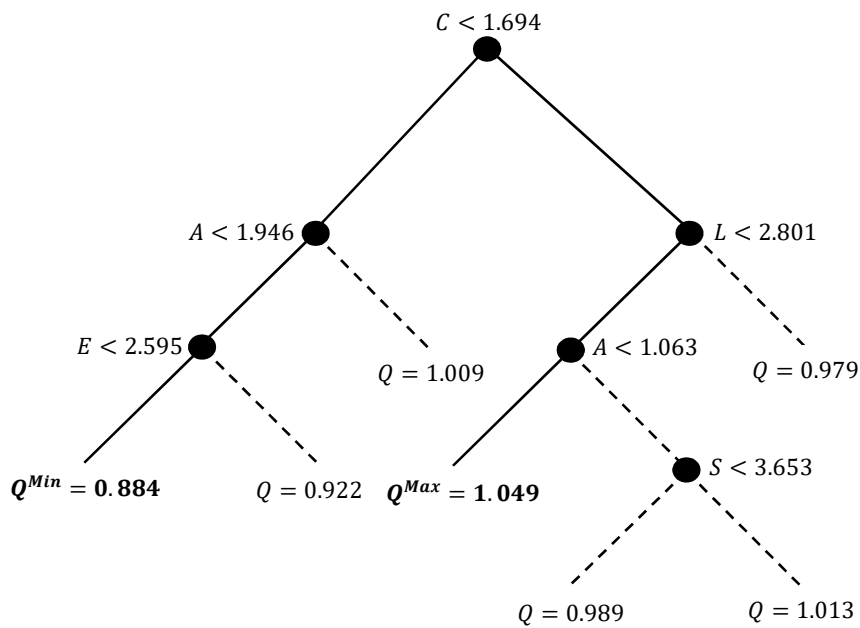

Fig. 4.b

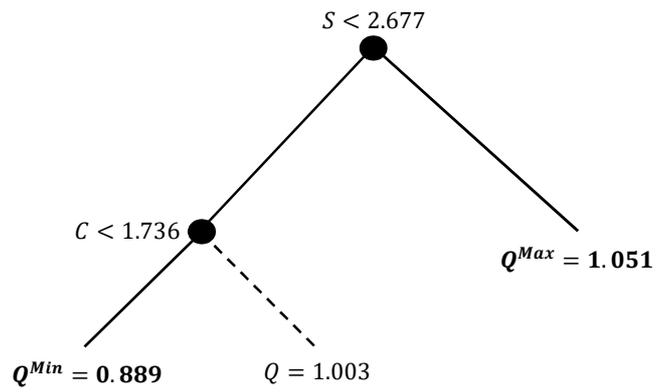

Fig. 4.c

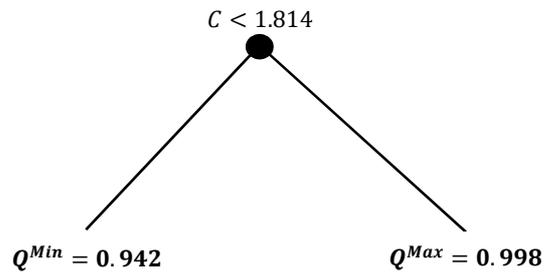

Fig. 4.d

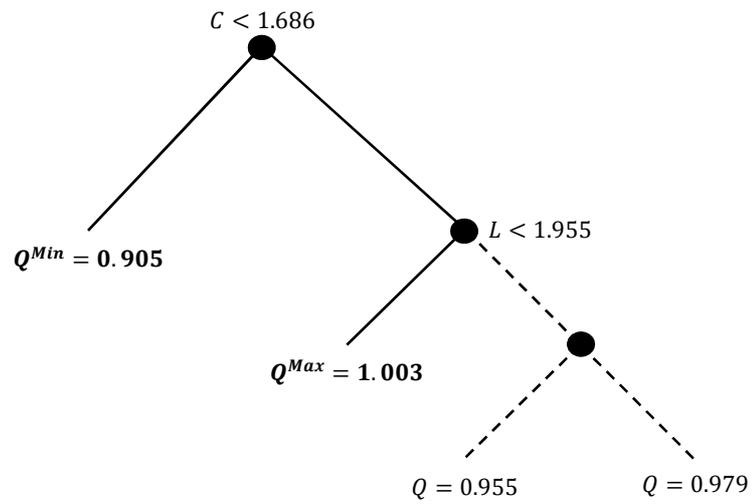

Fig. 4.e

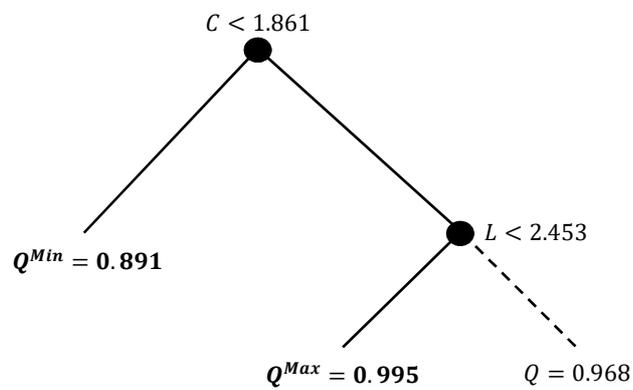

Insert Fig. 5.a
(see below)

**(a) Sample period, 2005–2016**

Insert Fig. 5.b
(see below)

Insert Fig. 5.c
(see below)

**(b) Pre-crisis period, 2005–2007**

**(c) Financial crisis, 2008–2009**

Insert Fig. 5.d
(see below)

Insert Fig. 5.e
(see below)

**(d) Sovereign debt crisis, 2010–2013**

**(e) Post-crisis, 2014–2016**

**Fig. 5. PIGS countries.** This set of figures show the trees generated by CART for the PIGS countries in 2005–2016 and the pre-crisis, financial crisis, sovereign debt crises, and post-crisis periods. The variables $C$, $A$, $E$, $L$, and $S$ represent the proxies selected by the RF algorithm to capture capital, asset, earnings, liquidity, and systematic risks, respectively, and $Q$ measures charter value. The variables capturing types of risk are rescaled using a scale of one to five. At any given node, the left branch includes banks satisfying the inequality shown in that node. The value of $Q$ at a terminal node equals the mean charter value of the group of banks in that terminal node. For the subperiods 2005–2007, 2008–2009, 2010–2013, and 2014–2016, the trees shown are generated with no stopping rule, that is, we do not request terminal nodes to include at least 30 banks. $Q^{Min}$ and $Q^{Max}$ stand for the minimum and maximum, respectively, of all the mean charter values corresponding to the sets of banks at the terminal nodes.

Fig. 5.a

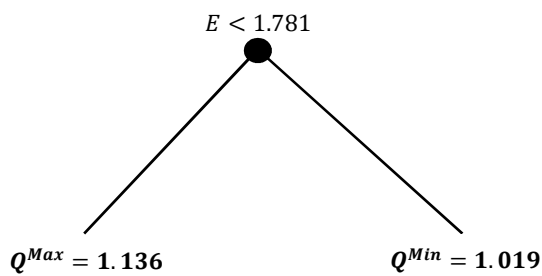

Fig. 5.b

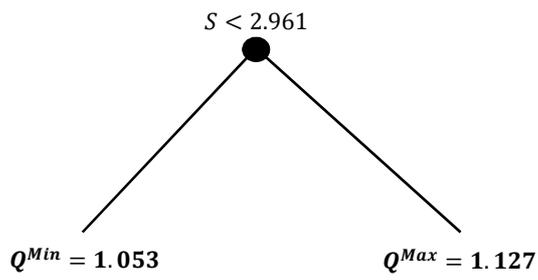

Fig. 5.c

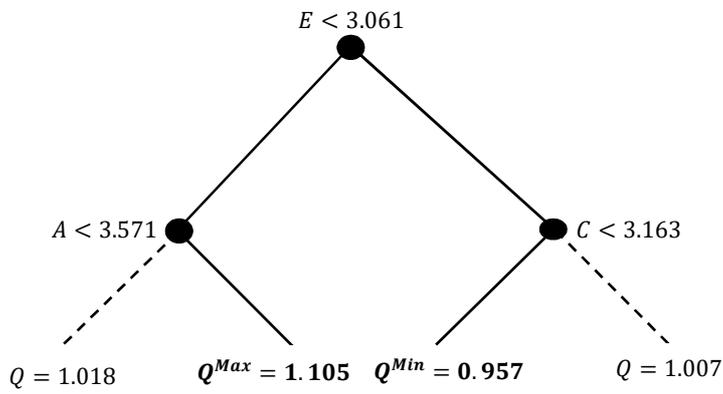

Fig. 5.d

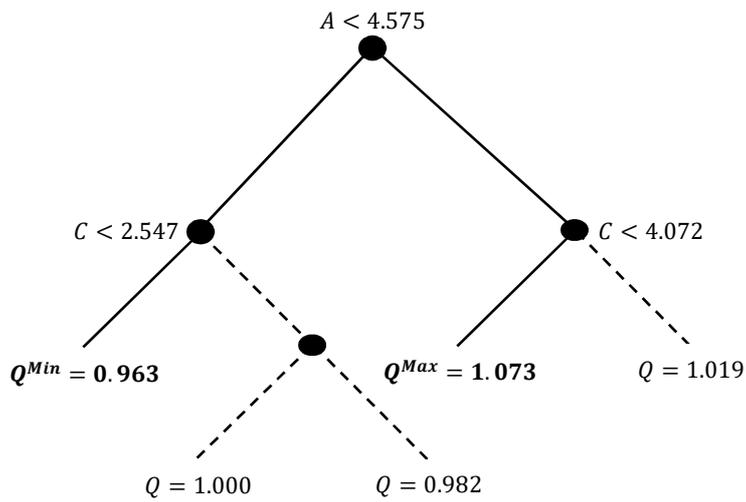

Fig. 5.e

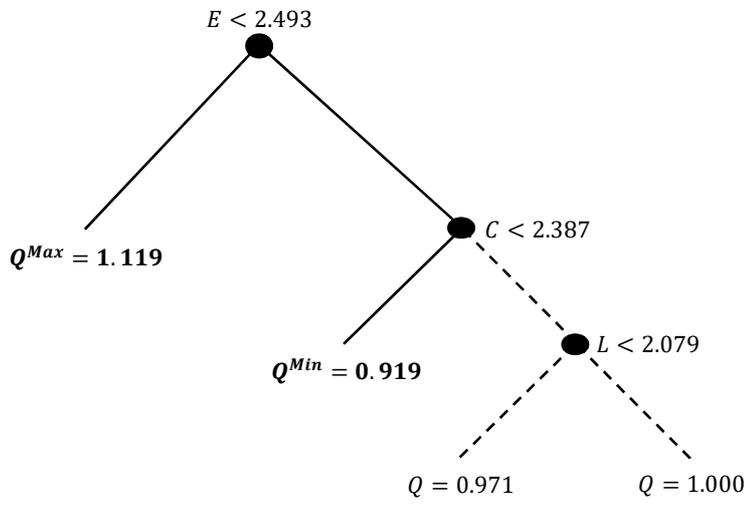

Insert Fig. 6.a
(see below)

**(a) Large banks**

Insert Fig. 6.b
(see below)

**(b) Small banks**

**Fig. 6. Large and small banks.** This set of figures show the trees generated by CART for large and small banks in the sample period. The variables $C$, $E$, $L$, and $S$ represent the proxies selected by the RF algorithm to capture capital, earnings, liquidity, and systematic risks, respectively, and $Q$ measures charter value. The variables capturing types of risk are rescaled using a scale of one to five. At any given node, the left branch includes banks satisfying the inequality shown in that node. The value of $Q$ at a terminal node equals the mean charter value of the group of banks in that terminal node. $Q^{Min}$ and $Q^{Max}$ stand for the minimum and maximum, respectively, of all the mean charter values corresponding to the sets of banks at the terminal nodes.

Fig. 6.a

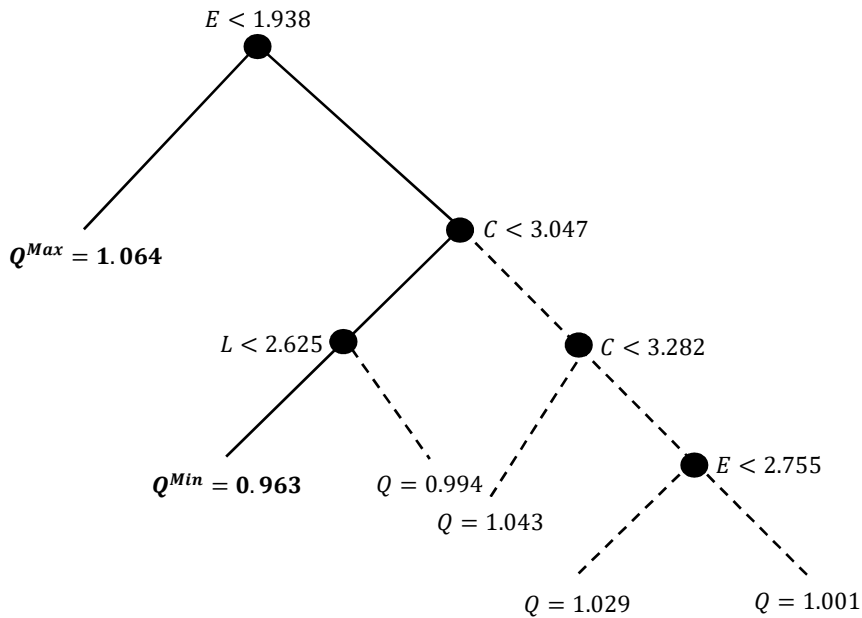

Fig. 6.b

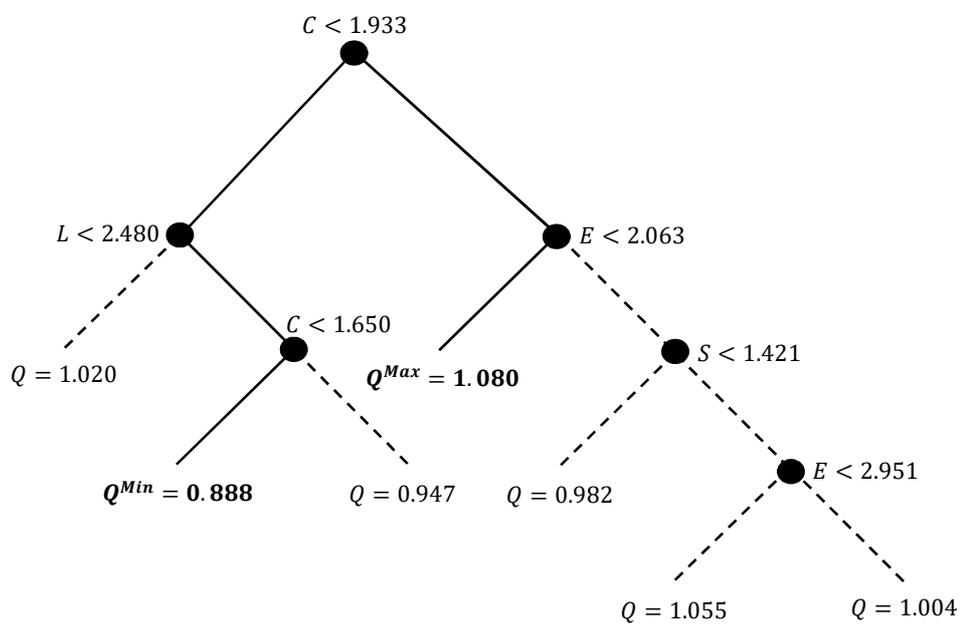